\documentclass{iopart}

\usepackage{cite}

\newcommand{\beq}{\begin{equation}}
\newcommand{\eeq}{\end{equation}}
\newcommand{\bea}{\begin{eqnarray}}
\newcommand{\eea}{\end{eqnarray}}
\usepackage{graphicx}
\begin{document}
\title{Structures of quantum 2D electron--hole plasmas}
\author{
V S Filinov$^{1}$, M Bonitz$^{2}$, H Fehske$^{3}$, P R
Levashov$^{1}$, V E Fortov$^{1}$}
\address{
$^1$Joint Institute for High Temperatures, Russian Academy of Sciences,\\
Izhorskaya 13 bldg 2, Moscow 125412, Russia\\
$^2$Christian-Albrechts-Universit{\"a}t zu Kiel, Institut f\"ur
Theoretische Physik und Astrophysik, Leibnizstrasse 15, 24098
Kiel, Germany\\
$^3$Institut f\"ur Physik, Ernst-Moritz-Arndt-Universit{\"a}t
Greifswald, Felix-Hausdorff--Str. 6, D-17489 Greifswald, Germany}
\date{\today}

\begin{abstract}
We investigate structures of 2D quantum electron--hole (e-h) plasmas
by the direct path integral Monte Carlo method (PIMC) in a wide
range of temperature, density and hole-to-electron mass ratio. Our
simulation includes a region of appearance and decay of the bound
states (excitons and biexcitons), the Mott transition from the
neutral e-h plasma to metallic--like clusters, formation from
clusters the hexatic-like liquid and formation of the crystal-like
lattice.
\end{abstract}


\section{Introduction}

Strongly correlated two-dimensional quantum Coulomb systems are the
subject of intensive discussions.
In particular, it is known that the competition between
electrostatic and kinetic energy in an electron gas may be the
reason of an unusual phase diagram of a 2D system of electrons. The
liquid state of such system is stable when the kinetic energy
dominates while the electrostatically favored ``Wigner'' triangular
crystal is stable in the opposite case. If there is a strong
competition between these two
kinds of energy, different situations are possible. 
The question under discussion is the existence of the intermediate
anisotropic liquid phase (hexatic) under melting of crystal into
isotropic liquid. Moreover, the physical mechanism of melting can be
influenced by the interaction with substrate and defects. 
Currently all mentioned phenomena can be extensively investigated by
a consistent first-principle numerical simulation, and in this brief
paper we present the most interesting results of our numerical
experiments by the direct PIMC method.

\section{Path integral Monte Carlo approach}

Let us consider a two-component neutral e-h plasma. Thermodynamic
properties of such plasma are defined by the partition function $Z$,
which for the case of $N_e$ electrons and $N_h$ holes ($N_e=N_h$),
is given by
$Z(N_e,N_h,V,\beta) = Q(N_e,N_h,\beta)/N_e!N_h$,
with $Q(N_e,N_h,\beta) = \sum_{\sigma}\int_V dr \,dq \,\rho(q,
r,\sigma;\beta)$, where $\beta=1/k_B T$,
$\sigma=(\sigma_e,\sigma_h)$, $r$, $\sigma_e$ denotes the space and
spin electron coordinates, while $q,\sigma_h$ denotes the space and
spin hole coordinates. These expressions are valid for 1D, 2D and 3D
cases.
The pair distribution functions 
for a binary mixture of quantum electrons and holes can be written
in the form 
$g_{ab}(R_1-R_2) = g_{ab}(R_1,R_2) =
\sum_{\sigma} \int_{V} \, dq \,dr \, \,\delta(R_1-Q_1^{a})\,
\delta(R_2-Q_2^{b}) \rho(q,r, \sigma, \beta)/Q(N_e,N_h,\beta)$,
 where $a$ and $b$ label the particle species, i.e. $a,b=e$
(electrons), $h$ (holes) and $Q,R$ denote the two-dimensional
vectors of coordinates, $Q^e_{1,2}=r_{1,2}$ and $Q^h_{1,2}=q_{1,2}$.
The exact density matrix $\rho(q,r,\sigma,\beta)$ of a quantum
system for low temperature and high density is in general not known
but can be constructed using a path integral representation
\cite{zamalin,PhRv2007}. Thus we take into account interaction,
exchange (through permutation operators) and spin effects both for
electrons and holes. This procedure gives the expression for the
density matrix as a multiple integral which is well suited for
efficient numerical evaluation by Monte Carlo techniques, e.g.
\cite{zamalin}. We mention here that for all obtained results, the
maximum statistical and systematic errors is not more than
$5\%$. 

\section{Simulation results}\label{simulations}

We are interested in strong Coulomb correlation effects such as
bound state (excitons, bi-excitons, many particle clusters), their
transformation by the surrounding plasma and their eventual breakup
at high densities (Mott effect). Beyond the Mott density, we expect
the possibility of the hole crystallization if the hole mass is
sufficiently large \cite{Bonitz_PRL05}. Below, the density of the
two-component plasma is characterized by the Brueckner parameter
$r_s$ defined as the ratio of the mean distance between particles
$d=[1/\pi (n_e+n_h)]^{1/2}$ and the 3D exciton Bohr radius $a_B$,
where $n_e$ and $n_h$ are the electron and hole 2D densities. The
dimensionless temperature will be presented as a ratio of the
temperature and the 3D electron-hole binding energy (Rydberg), which
includes the reduced effective mass and dielectric constant.
\begin{figure}[htb]
\centering
\includegraphics[width=0.40\columnwidth]{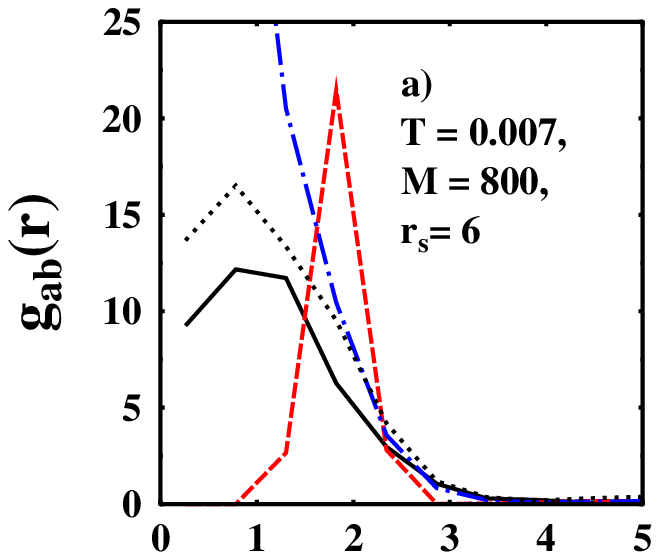}
\hspace{0.5cm}
\includegraphics[width=0.305\columnwidth,trim=0cm -1.05cm 0cm 0cm]{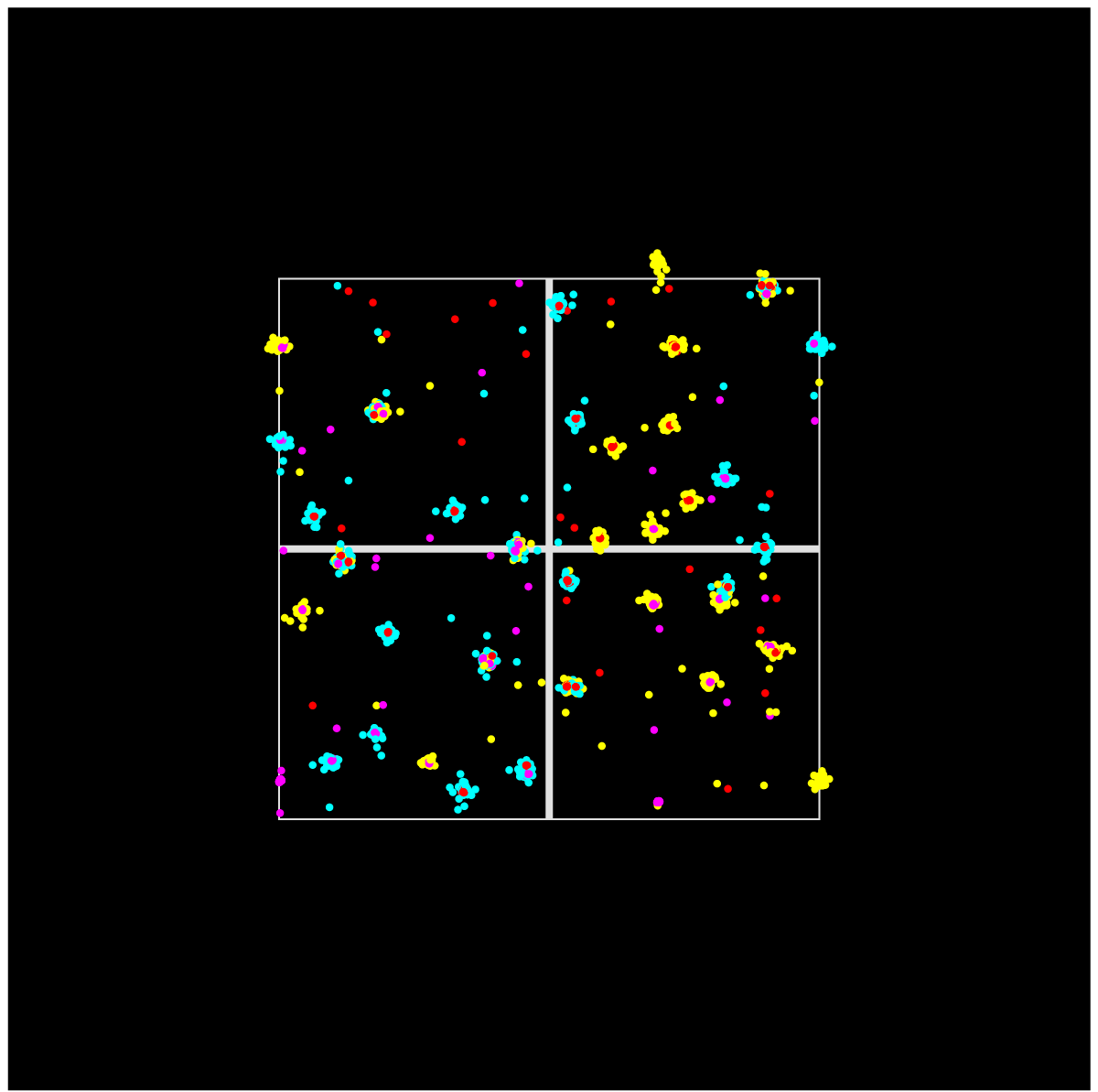}
\\
\hspace{1.5mm}
\includegraphics[width=0.40\columnwidth]{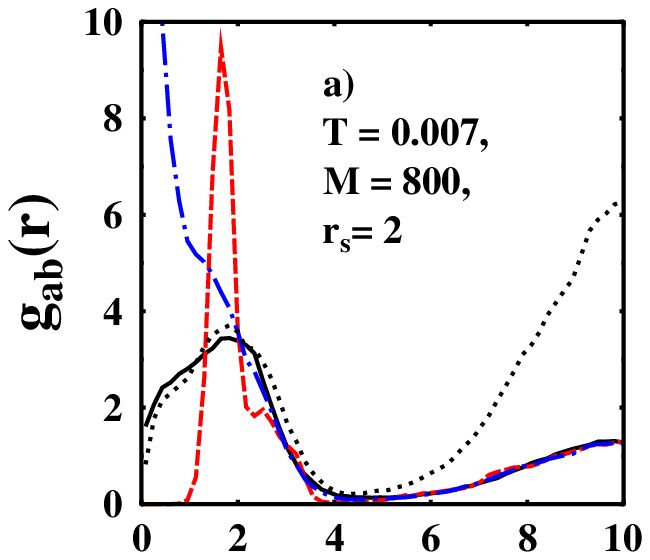}
\hspace{0.3cm}
\includegraphics[width=0.31\columnwidth,trim=0cm -1.05cm 0cm 0cm]{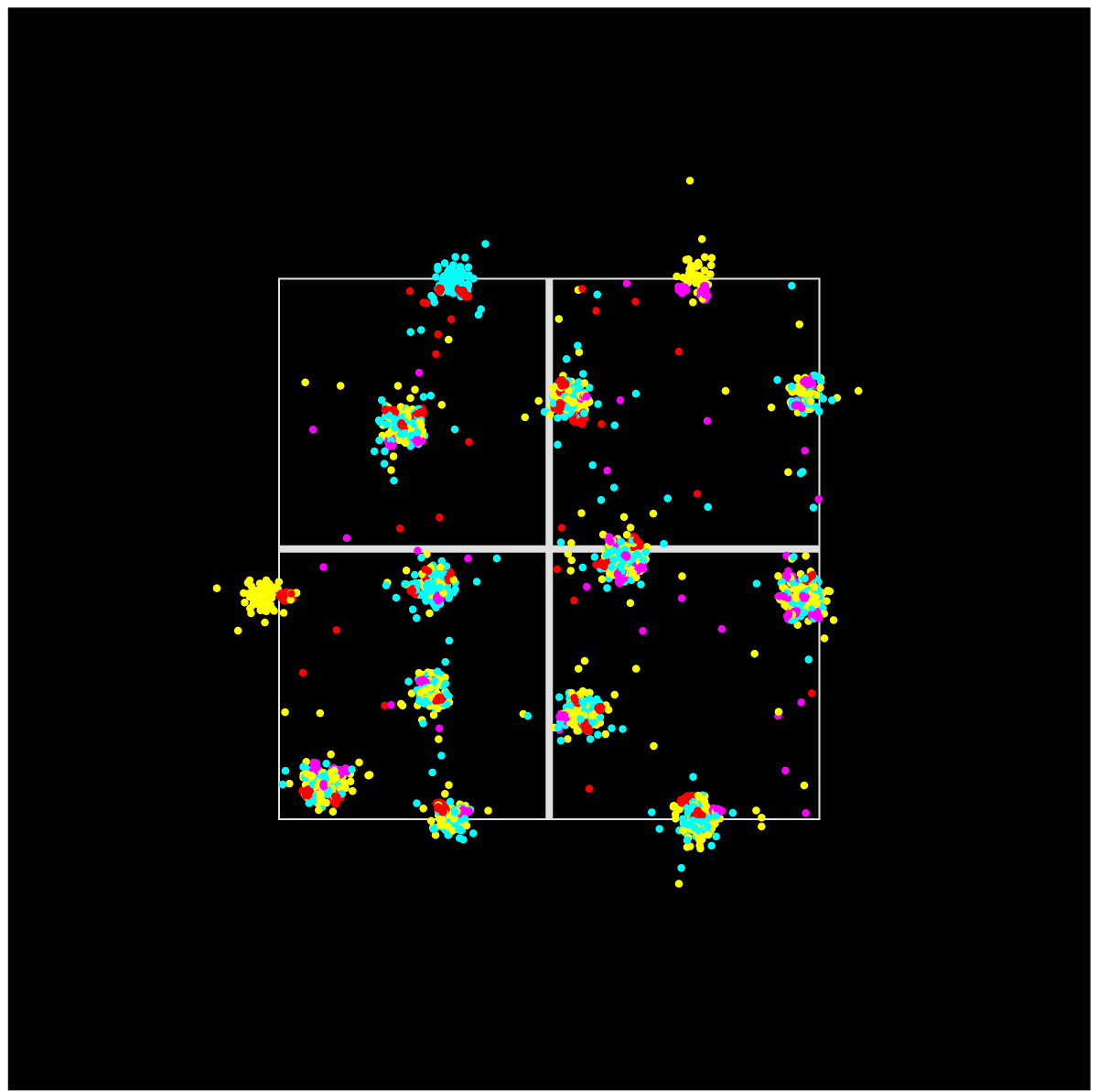}
\caption{(Color online) Pair distribution functions (left column)
and snapshots of the Monte Carlo cell (right column). Left column:
$g_{ee}$ (black solid line), $g_{hh}$ (red dashed line), $g_{eh}$
(blue dot-dashed line) at temperature $T/Ry = 0.007$, the hole to
electron mass ratio $M=800$ and densities related to the Brueckner
parameter $r_s=6$ and $r_s=2$. Right column: red and magenta
clouds---holes, yellow and cyan clouds---electrons with the opposite
spin directions.}
 \label{f1}
\end{figure}

We analyze some spacial distribution functions and related
spin-resolved typical ``snapshots'' of the e-h state in the
simulation box for different particle densities, temperatures and
hole-to-electron mass ratios $M$.
According to the path integral representation of the density matrix,
each electron and hole is represented by several tens of points
(``beads''). The spatial distribution of the beads of each quantum
particle is proportional to its spatial probability distribution.
\Fref{f1} shows that, the typical size of the cloud of beads for
electrons is several times larger than the one for the heavy holes.
At low temperature and low ($r_s=6$) and middle ($r_s=2$) densities
 practically all holes are closely covered by electron
beads. From a physical point of view this means that electrons and
holes form bound states, i.e. excitons, bi-excitons and many
particle clusters. The existence of the bound states is also
supported by the behavior of the pair distribution functions,
exhibiting pronounced maxima at the distances of about half and a
bit more of the Bohr radius. We note that raising the temperature at
a fixed density leads to a temperature-induced ionization of the
bound states. As a result we found a substantial number of free
electrons and holes.

From \fref{f1} it follows that the growth of density results in the
increase of the number of particles in clusters. The structural
analysis of large many-particle clusters show the hexagonal ordering
of heavy holes inside liquid-like clusters
(lower right panel). In this panel there are only two clusters: one
is in the center of the Monte Carlo cell, while the second one is
divided in four parts (in each corner) due to the periodic boundary
conditions of the Monte Carlo cell. Here besides the inner normal
hexagonal structure the holes due to the strong Coulomb repulsion
form the filament-like structures of the clouds of beads at the
bounds of clusters. The filament-like structures are fully 2D
topological effect as in 3D case the holes in analogous many
particles clusters have normal liquid-like ordering
\cite{PhRv2007,PSS}.

\begin{figure}[htb]
\centering
\includegraphics[width=0.40\columnwidth]{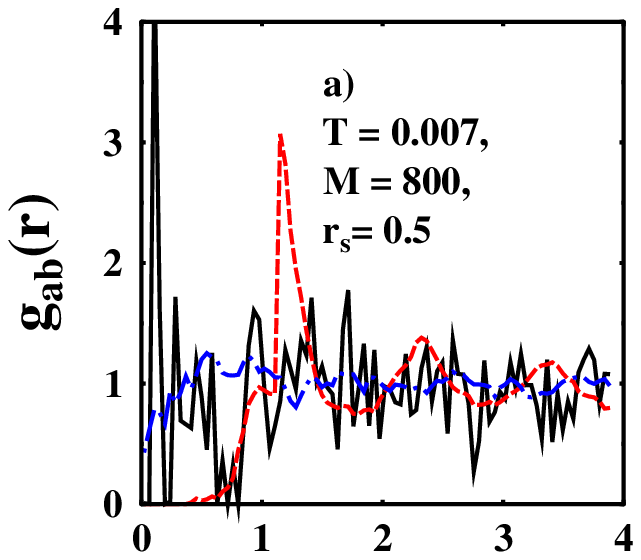}
\hspace{0.5cm}
\includegraphics[width=0.315\columnwidth,trim=0cm -1.1cm 0cm 0cm]{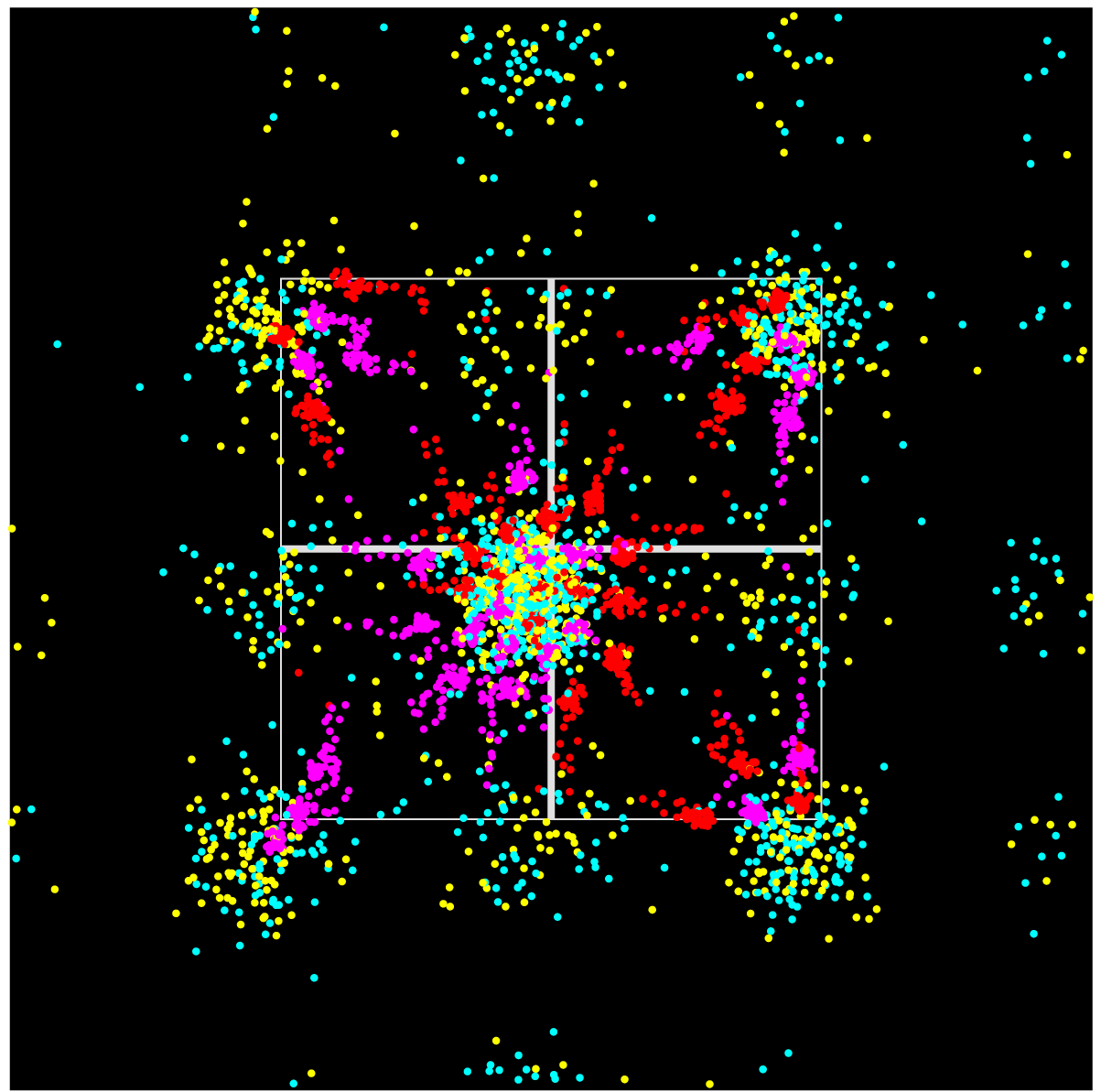}
\centering
\includegraphics[width=0.40\columnwidth]{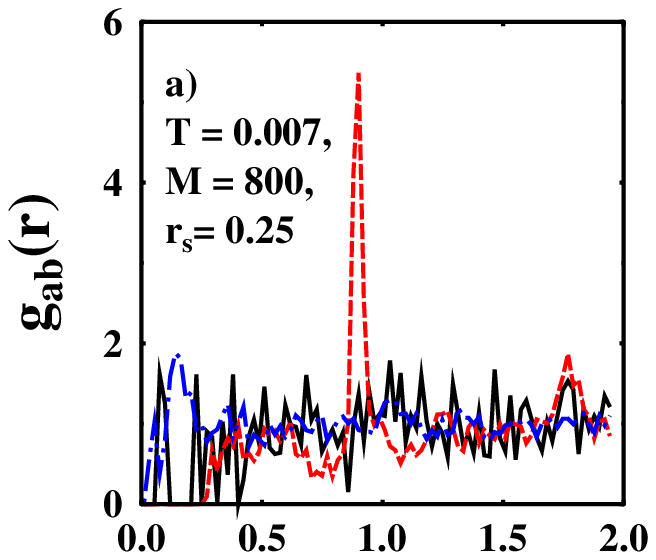}
\hspace{0.5cm}
\includegraphics[width=0.315\columnwidth,trim=0cm -1cm 0cm 0cm]{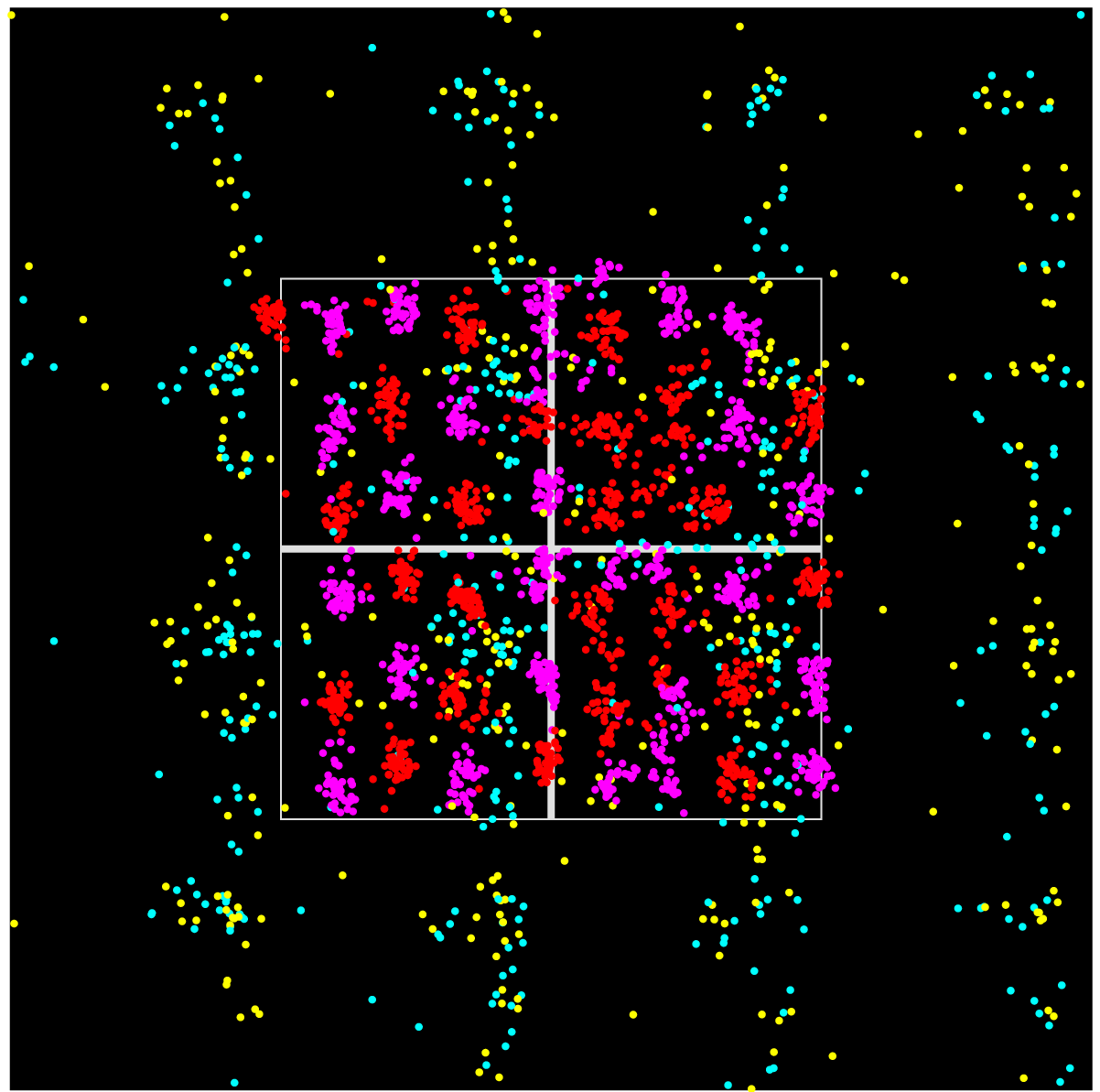}
\caption{(Color online) Pair distribution functions (left column)
and snapshots of the Monte Carlo cell (right column) at temperature
$T/Ry = 0.007$, the hole-to-electron mass ratio $M=800$ and
densities related to the Brueckner parameter $r_s=0.5$ and
$r_s=0.25$. Notations are the same as in \fref{f1}.}
 \label{f2}
\end{figure}

If the particle density is high enough ($r_s=0.25$) the electron
wavelength becomes larger than the mean inter-particle distance $d$
and even larger than the size of the Monte Carlo cell which is seen
by the large extension of the clouds of electron beads. For $r_s \le
0.5$ clusters become unstable because two electrons bounded to
neighboring holes start to overlap allowing for electron tunneling
from one cluster to the other (Mott effect). Since the hole
wave-length is significantly smaller than the electron wavelength,
it may still be smaller than $d$ and the structure of the hole beads
resembles a liquid-like or crystal-like state. If the hole mass
exceeds a critical value, the holes may even form a crystal-like
structure \cite{Bonitz_PRL05} (lower right panel in \fref{f2}). Here
the holes form a crystal-like structure, while the electron density
demonstrates the Bloch oscillations. At this very high density the
type of the hole 'crystal' is influenced by the boundary conditions
of the Monte Carlo cell (finite size effect). The detailed analysis
of the type of the crystal-like structure is possible for a much
more bigger number of particles in the Monte Carlo cell.

\section{Conclusion}\label{summary}

In this paper we have presented a computer simulation analysis of
strong Coulomb correlations in dense two-dimensional two-component
quantum plasmas at low temperatures. In particular, the formation
and dissociation of bound states, such as excitons, bi-excitons and
many particle clusters, is analyzed and the density-temperature
regions of their occurrence is identified. We have shown that, above
the Mott point, two-component plasmas with large mass anisotropy
show interesting Coulomb correlation phenomena: with increasing
density holes can undergo a phase transition to a Coulomb
hexatic-like liquid and to a Wigner crystal which are embedded into
a degenerate electron gas. The crystal-like formation in a
two-component plasma is possible for large enough hole-to-electron
mass ratios \cite{Bonitz_PRL05}.

\section{Acknowledgments}
V. Filinov acknowledges the hospitality of the Institut f\"ur
Theoretische Physik und Astrophysik of the
Christian-Albrechts-Universit{\"a}t zu Kiel and Institut f\"ur
Physik, Ernst-Moritz-Arndt-Universit{\"a}t zu Greifswald.

\end{document}